\documentclass{JHEP3} 

\usepackage{epsfig}
\usepackage{graphicx}

\usepackage{cite}
\usepackage[normalem]{ulem}
\usepackage{graphics}
\usepackage{amsmath}
\usepackage{amsthm}
\usepackage{amssymb}
\usepackage{enumerate}
\usepackage{bm,longtable,enumerate}
\usepackage{amsmath,amssymb}

\newcommand{\ol}{\overline}

\def\tr{\mathop{\rm tr}\nolimits}

\newcommand{\AD}[1]{$\ol{\mbox{D~\,}}\!\!\!$#1}

\newcommand{\eqn}{\begin{eqnarray}}
\newcommand{\eqnx}{\end{eqnarray}}

\def\beq{\begin{equation}}
\def\eeq{\end{equation}}
\def\beqa{\begin{eqnarray}}
\def\eeqa{\end{eqnarray}}
\def\ss{\scriptscriptstyle}

\def\matt[#1,#2,#3,#4]{\left(%
\begin{array}{cc} #1 & #2 \\ #3 & #4 \end{array} \right)}

\def\v2#1{\vv2[#1]}
\def\vv2[#1,#2]{\left(%
{#1 \atop #2}\right)}

\title{Form factors of vector and axial-vector mesons in holographic D4-D8 model}

\author{C.  A.  Ballon Bayona$^{\dag}$, Henrique Boschi-Filho$^{\ddag,a}$,
Nelson R. F. Braga$^{\ddag,b}$and Marcus A. C. Torres$^{\ddag,c}$
\\
$^\dag$Centro Brasileiro de Pesquisas F\'{i}sicas, Rua Dr. Xavier Sigaud 150,\\
Urca, 22290-180, Rio de Janeiro, RJ, Brazil\\
e-mail: ballon@cbpf.br
\\
\\
$^{\ddag}$Instituto de F\'{i}sica, Universidade Federal do Rio de Janeiro,\\
Caixa Postal 68528, RJ 21941-972, Brazil\\
$^{a}$e-mail: boschi@if.ufrj.br;
$^{b}$e-mail: braga@if.ufrj.br;
$^{c}$e-mail: mtorres@if.ufrj.br}

\abstract{ We calculate  elastic and non-elastic electromagnetic form factors
for  vector and axial-vector mesons in the holographic D4-D8 brane model.
We obtain the mass spectrum and Regge trajectories for these particles. 
From the elastic form factors we extract the electric radius, the magnetic and quadrupole moments. 
Form factors for transverse and longitudinal polarizations are also obtained.
We find superconvergence sum rules for the vector and axial-vector meson couplings 
that determine the asymptotic behavior of the form factors at large momentum transfer.
Our results show good agreement with  other holographic models and QCD. }

\preprint{}

\keywords{Gauge-gravity correspondence, AdS-CFT Correspondence}
 
\begin{document}

\section{ Introduction }

Recently a holographic model for large $N_c$ QCD with massless quarks at strong coupling was proposed 
\cite{Sakai:2004cn,Sakai:2005yt} (see also \cite{Hata:2007mb,Hashimoto:2008zw,Hashimoto:2009ys}). This model consists on the intersection of $N_c$ D4-branes and $N_f$  D8-\AD8 pairs of branes in type IIA string theory in the limit $N_f \ll N_c$. The numbers $N_c$ and $N_f$ are interpreted as the color and flavor numbers of strongly coupled QCD like theory. A remarkable feature of the D4-D8 brane model is the holographic description of chiral symmetry breaking $U(N_f)_L \times U(N_f)_R \to U(N_f)$ from the merging of the D8-\AD8 branes.   

One important property of the D4-D8 brane model is the explicit realization of vector meson dominance 
\cite{GellMann:1961tg,Sakurai_69} which states that the interaction of a hadron with a photon is mediated by vector mesons.
Therefore, the calculation of hadronic form factors involve a sum over all intermediate vector mesons that, in this model, correspond to the Kaluza Klein modes of an extra compact dimension. 
This property makes the D4-D8 model an important tool to describe the electromagnetic interaction of hadrons.  
In particular, the results for the pion form factor obtained in ref.\cite{Sakai:2005yt} present  good agreement with available experimental data. 

In this article we use the D4-D8 brane model to study the electromagnetic interactions of vector and axial-vector mesons. First we obtain numerically the wave functions that arise in the Kaluza Klein expansion in the compact extra dimension. The associated eigenvalues give us the meson mass spectra and Regge trajectories.
We calculate the relevant decay constants and the couplings of three-meson vertices.  Using these results we 
obtain the  electromagnetic elastic and transition form factors for vector and axial-vector mesons. 

We analyse the momentum dependence  of these form factors in the limit of large space-like momentum transfer.   
We find superconvergence sum rules for the vector and axial-vector meson coupling constants that determine the 
asymptotic behavior  of the form factors. These sum rules generalize a previous result for the $\rho$ meson obtained within the hard wall 
model \cite{Grigoryan:2007vg}.

From the elastic form factors, we extract the electric radius, the magnetic and quadru\-pole moments
for the $\rho$ and $a_1$ mesons. We calculate the elastic form factors $ F_{TT}$, $F_{LT}$,  $F_{LL}$ that arise from the decomposition of the initial and final meson polarizations into transversal and longitudinal components.  We find that the asymptotic behavior of these form factors in the D4-D8 model is in agreement with QCD predictions.

Electromagnetic form factors of vector mesons in other holographic models  have been studied in 
\cite{Hong:2003jm,Hong:2004sa,Grigoryan:2007vg,Grigoryan:2007my,Brodsky:2007hb}. 
For a review of mesons in holographic models see \cite{Erdmenger:2007cm}.
Form factors of baryons have been calculated recently in the D4-D8 brane model by  considering instanton configurations \cite{Hashimoto:2008zw,Hashimoto:2009ys}.

\section{D4-D8 Model}
 The Sakai-Sugimoto D4-D8 model is built adding $N_f$ pairs of \mbox{D8}  and \AD8 probe branes in the spacetime background formed by the presence of $N_c$ \mbox{D4} branes. Initially, the branes are set in the following spacetime configuration:

\begin{table}[!ht]
\begin{center}
\begin{tabular}{|r|c|c|c|c|c|c|c|c|c|c|}
\hline
& 0 & 1 & 2 & 3 & (4) & 5 & 6 & 7 & 8 & 9 \\
\hline
D4 & $\circ$ & $\circ$ & $\circ$ & $\circ$ & $\circ$ &&&&& \\
\hline
D8 - \AD8 & $\circ$ & $\circ$ & $\circ$ & $\circ$ &  & $\circ$ & $\circ$ & $\circ$ & $\circ$ & $\circ$ \\
\hline
\end{tabular}
\label{intersecao}
\end{center}
\end{table}

The brane worldvolumes are extended in the directions indicated by $\circ$. The four directions $x_0,\dots, x_3$ correspond to our four-dimensional Minkowski space-time and the direction $x_4$ is compactified in a circle. 

The low energy spectrum of open strings in the D4-D8 geometry includes gauge fields in the adjoint representation (arising from 4-4 strings) and quarks in the fundamental representation (coming from 4-8 strings) of the group $U(N_c)$. All these fields live in a 5d space-time with a compact coordinate  where fermionic anti-periodic conditions are imposed in order to  break supersymmetry.

Since the quarks D4-D8 and D4-\AD8 have different chirality we would expect the presence of chiral symmetry $U(N_f)_L\times U(N_f)_R$. However the background forces the branes D8 and \AD8  to be united in a single stack of $N_f$ D8 branes, breaking chiral symmetry $U(N_f)_L\times U(N_f)_R$ into a remaining $U(N_f)$. 

\subsection{The D4 background}

The D4-brane solution consists on the space-time metric 
\beq
ds^2 \,=\, \left (\frac{U}{R}\right)^{3/2}
\left[\eta_{\mu\nu}dx^\mu dx^\nu + f(U) d\tau^2\right]
+\left(\frac{R}{U}\right)^{3/2}
\left[\frac{dU^2}{f(U)}+U^2 d\Omega_4^2\right], 
\label{D4sol}
\eeq
with $f(U)=1-(U_{\ss{KK}}/U)^3$, and the dilaton and RR form 
\beq
e^\Phi \,=\, g_s \left(\frac{U}{R}\right)^{3/4} \qquad , \qquad F_4\,=\, \frac{3 N_c}{ 4 \pi } \epsilon_4 \, .
\eeq
The $x^\mu$ represent the 4d coordinates in Minkoswki spacetime with metric $\eta_{\mu \nu}$, while $\tau$ is the compactified $x_4$ direction, with period $\delta \tau$. The $x_5,\dots,x_9$ are written in spherical coordinates, with $U$ its radial direction. The constant $R$ is related to the string length $\sqrt{\alpha'}$ and string coupling $g_s$ by $R^3=\pi g_s N_c \alpha'^{3/2}$ while $\epsilon_4$ is the volume form. 

The radial coordinate has a minimum $U_{\ss{KK}}$ related to the  period of the $\tau$ coordinate by 
\beq
\delta \tau \,=\, \frac{4 \pi}{3} \frac{R^{3/2}}{U_{\ss{KK}}^{1/2}} \, , 
\eeq
which is obtained from the singularity avoidance condition. The Kaluza-Klein modes associated with the compact 
$\tau$ coordinate will thus have a natural mass scale 
\beq
M_{\ss{KK}}\,\equiv \, \frac{2 \pi}{\delta\tau} \,=\, \frac32 \frac{U_{\ss{KK}}^{1/2}}{R^{3/2}} \,.
\eeq

The effective four dimensional Yang-Mills coupling is related to the string coupling by 
\beq
g_{YM}^2 \,=\,2 \pi \sqrt{\alpha'} M_{\ss{KK}} g_s \,.
\eeq 

\subsection{The D8-brane solution}

The D8-brane is localized by imposing a relation between the radial coordinate $U$ and the compact coordinate $\tau$ in (\ref{D4sol}). The configuration $U=U(\tau)$ has to  minimize the Dirac-Born-Infeld action
\beqa
S_{D8} &\,=\,& - \mu_8\int d^4 x d\tau d^4 \Omega \, e^{-\Phi} \sqrt{- \det{ {\cal P}[G]_{AB} }} \nonumber \\
&\,\sim \,& \int d\tau \, U^4 \sqrt{ f(U) + \frac{R^3}{U^3} \frac{U'^2}{f(U)}}
\eeqa
where $\mu_8 = (2 \pi)^{-8} \alpha'^{-9/2}$ is the D8 brane tension, $ {\cal P}[G]_{AB}$ is the induced metric of the D8 brane, 
and $U'=dU/d\tau$. 
We are looking for a brane configuration symmetric in $\tau $ with a minimum value $ U^\ast $.
Then, the appropriate boundary conditions are  $U'(0)=0$ and $U(0) = U^\ast $. Since the action does not depend explicitly on $\tau$ we can easily find $U(\tau)$ as a solution to the energy conservation equation 
\beq
\left ( \frac{U^4 f(U)}{\sqrt{ f(U) + \frac{R^3}{U^3} \frac{U'^2}{f(U)}}} \right ) \,=\, U_\ast^4 \sqrt{f(U_\ast)} \, ,
\eeq
After imposing the Sakai-Sugimoto condition $U_\ast=U_{\ss{KK}}$ the D8-brane equation reduces to 
\beqa
\frac{d\tau}{d U} \,=\, 
\left\{
\begin{array}{ll} 0  \qquad  {\rm for} \quad U>U_{\ss{KK}}  \\
        \pm \infty   \qquad {\rm for} \quad U=U_{\ss{KK}}  \end{array}
         \right. \, ,  \, \,
\eeqa
which has a box-type solution consisting on two parallel lines $\tau(u)\,=\,\pm \delta \tau / 4$ plus a perpendicular line $U\,=\,U_{\ss{KK}}$. In the dual gauge theory, the condition $U_\ast=U_{\ss{KK}}$ can be interpreted as a zero mass condition for the quarks.  It is convenient to define the coordinates 
\beq
r \,=\, U_{\ss{KK}} \sqrt{ \frac{U^3}{U_{\ss KK}^3}-1} \quad , \quad \theta \,=\, \frac{2\pi}{\delta \tau} \tau  \quad , \quad 
( y , z) \,=\, (r \cos{\theta} , r \sin{\theta}) \,, 
\eeq 
so that the D8-brane solution is simply $y\,=\,0$ with the $z$ coordinate running from $-\infty$ to $\infty$. The D8-brane metric then can be written as \cite{Sakai:2004cn}
\beq
ds^2_{\ss D8} \,=\, \left(\frac{U_z}{R}\right)^{3/2}
\eta_{\mu\nu} dx^\mu dx^\nu \,+\, \frac49 \left(\frac{R}
{U_z} \right)^{3/2}\frac{U_{\ss{KK}}}
{U_z} dz^2 \,+\, R^{3/2}U_z^{1/2} d\Omega_4^2 \, , \label{D8branemetricfinal}
\eeq
with 
\beq
U_z\,=\,U_{\ss{KK}} \left( 1 + \frac{z^2}{U^2_{\ss KK }} \right)^{1/3} \,.
\eeq
Finally, the probe limit condition $N_f<<N_c$ assures that the D4 background deformation by the D8/\AD8 branes can be neglected. 
This condition can be interpreted in the dual theory as a quenching limit, where the quark loops are singled out. 
 
 \section{Vector and axial-vector mesons }

In the D4-D8 model, mesons of the dual gauge theory emerge as states of open strings  connecting the D8-\AD8 branes. 
These states correspond to fluctuations of the D8 branes solutions in the D4 background. 
In particular, vector and axial-vector mesons are described by  $U(N_f)$ gauge field fluctuations. 
The dynamics of these fluctuations is given by the action 
\beq
S_{D8}\,=\, (\pi \alpha')^2 \mu_8 \int d^4 x  dz d^4 \Omega e^{-\Phi} \sqrt{ - \det g }\, \tr( F^{MN}F_{MN})  \,+\,  S_{CS} \, , \label{gaugefieldaction}
\eeq
where $F_{MN}=\partial_M A_N - \partial_N A_M + [A_M , A_N ]$ is the field strength with $A_M=(A_\mu, A_z , A_\alpha)$ 
and $g_{MN}$ is the D8-brane metric given in eq. (\ref{D8branemetricfinal}). 
The first term in the action (\ref{gaugefieldaction}) arises from the $\alpha'$ expansion of the non-abelian 
Dirac Born Infeld action (see for instance \cite{Tseytlin:1997csa}). 
The second term is the Chern-Simons action which is not relevant for the present discussion. 
We will set $A_\alpha=0$ and assume that the other gauge field components do not depend on the $S^4$ coordinates. 
Then, we obtain a five dimensional effective action for gauge field fluctuations 
\beq
S_{eff}\,=\, \kappa \int d^4 x \int d \tilde z \, \tr  \, \left [ \frac{1}{2} (K (\tilde z))^{-1/3} \eta^{\mu \lambda} \eta^{\nu \rho} F_{\lambda \rho} F_{\mu \nu} + M^2_{\ss{KK}} K(\tilde z) \eta^{\mu \nu} F_{\mu \tilde z} F_{\nu \tilde z}\right ] \label{effectivegaugeaction}
\eeq
where we introduced the dimensionless variable $\tilde z=z/U_{\ss{KK}}$ and  
\beq
K(\tilde z) \, \equiv \, 1 + \tilde z^2 \quad , \quad \kappa \,=\, \frac{g_{YM}^2 N_c^2}{ 216 \pi^3} \,.
\eeq
In order to obtain a four dimensional effective action with external gauge fields we choose the gauge condition $A_{\tilde z} = 0$ 
and expand $A_\mu$ in the following way
\cite{Sakai:2005yt} :
\beqa
A_{\mu} (x,\tilde z) &\,=\,&  \hat {\cal V}_\mu (x)  + \hat {\cal A}_\mu (x)  \psi_0 (\tilde z)  \,+\, \sum_{n=1}^{\infty} v_\mu^n (x) \psi_{2n-1} (\tilde z) \,+\,  \sum_{n=1}^{\infty} a_\mu^n (x) \psi_{2n} (\tilde z)    \quad , \quad \label{gaugefieldexpansion}
\eeqa
where $\psi_0 (\tilde z)\equiv (2/\pi ) \arctan{\tilde z}$ and 
\beqa
\hat {\cal V}_\mu (x)&\,=\,& \frac12  e^{- \frac{ i \Pi(x)}{f_\pi}}\left[ A_{L\mu}(x) + \partial_\mu \right] e^{\frac{ i \Pi(x)}{f_\pi}} \,+\, \frac12  e^{ \frac{ i \Pi(x)}{f_\pi}}\left[ A_{R \mu} (x) + \partial_\mu \right] e^{\frac{- i \Pi(x)}{f_\pi}}\nonumber \\
\hat {\cal A}_\mu (x) &\,=\,& \frac12  e^{- \frac{ i \Pi(x)}{f_\pi}}\left[ A_{L\mu} (x)+ \partial_\mu \right] e^{\frac{ i \Pi(x)}{f_\pi}} \,-\, \frac12  e^{ \frac{ i \Pi(x)}{f_\pi}}\left[ A_{R \mu} (x) + \partial_\mu \right] e^{\frac{- i \Pi(x)}{f_\pi}} \,.
\eeqa
The field $\Pi(x)$ is interpreted as the pion field, the fields $A_{L\mu}(x)$ and $A_{R\mu}(x)$ represent external gauge fields while the fields $v_\mu^n (x)$ and $a_\mu^n (x)$ are related to  vector and axial-vector mesons. 
Substituting (\ref{gaugefieldexpansion}) in (\ref{effectivegaugeaction}) and imposing the conditions 
\beqa
\kappa \int d \tilde z (K( \tilde z))^{-1/3} \psi_r (\tilde z) \psi_s (\tilde z) &\,=\,&  \delta_{rs} \, \, ,  \label{vectormesonnorm} \\
- (K(\tilde z) )^{1/3} \partial_{\tilde z} \left[ K(\tilde z) \partial_{\tilde z} \psi_r (\tilde z) \right] &\,=\,& \lambda_r \,  
\psi_r (\tilde z) \, \label{vectormesoneq},
\eeqa
where $r,s$ are positive integers, we obtain a 4d effective lagrangian where four dimensional vector and axial-vector 
mesons emerge as modes of the 5d flavor gauge fields after integrating out the $\tilde z$ direction. 
 The condition (\ref{vectormesonnorm}) is a usual orthonormality condition for the wave functions while the condition (\ref{vectormesoneq}) consists on an infinite set of equations of motion for the $\psi_r (\tilde z)$ modes.  

The 4d effective lagrangian after convenient field redefinitions (and apart from divergent terms) is 
\beqa
{\cal L}^{4d}_{eff} &\,=\,& \frac12 \tr \left(\partial_\mu \tilde v_\nu^n - \partial_\nu \tilde v_\mu^n \right)^2 + \frac12 \tr  \left(\partial_\mu \tilde a_\nu^n - \partial_\nu \tilde a_\mu^n \right)^2 + \tr  \left(i \partial_\mu \Pi + f_\pi {\cal A}_\mu \right)^2 \nonumber \\
&\,+\,& M_{v^n}^2 \tr  \left(\tilde v_\mu^n - \frac{g_{v^n}}{M_{v^n}^2} {\cal V}_\mu \right)^2 + M_{a^n}^2 \tr  \left(\tilde a_\mu^n - \frac{g_{a^n}}{M_{a^n}^2} {\cal A}_\mu \right)^2 \,+\, \sum_{j \ge 3} {\cal L}_j \,  \label{fourdimensionalefflag}
\eeqa
where ${\cal L}_j$ represent the interaction terms of order $j$ in the fields. 
The field redefinitions and constants are
\beqa
\tilde v_\mu^n &\,=\,& v_\mu^n + \frac{g_{v^n}}{M_{v^n}^2} {\cal V}_\mu  \quad \, \, \, \, , \quad 
\tilde a_\mu^n \,=\, a_\mu^n + \frac{g_{a^n}}{M_{a^n}^2} {\cal A}_\mu  \, ,  \\
 {\cal V}_\mu &\,=\,& \frac12 (A_{L\mu} + A_{R\mu})\quad , \quad {\cal A}_\mu \,=\, \frac12 (A_{L\mu} - A_{R\mu}) \, \\
M_{v^n}^2 &\,=\,& \lambda_{2n-1} M^2_{\ss{KK}} \qquad \, \, , \quad M_{a^n}^2 \,=\, \lambda_{2n} M^2_{\ss{KK}} \, , \label{masses}\\
 g_{v^n} &\,=\,& \kappa  \, M_{v^n}^2 \int d \tilde z \,K(\tilde z)^{-1/3} \psi_{2n-1}(\tilde z) \, , \label{decayvn}\\
g_{a^n} &\,=\,& \kappa \, M_{a^n}^2  \int d \tilde z \, K(\tilde z)^{-1/3} \psi_{2n}(\tilde z) \psi_0(\tilde z) \label{decayan} \, , 
\eeqa

The effective lagrangian (\ref{fourdimensionalefflag}) contains massive vector mesons $\tilde v_\mu^n$, $\tilde a_\mu^n$ and a massless pion field $\Pi$. The constant $g_{v^n}$ is the coupling between a  vector meson $\tilde v_\mu^n$ and an external $U(1)$ field ${\cal V}_\mu$ representing the photon. 
This is the only interaction between photons and mesons in the D4-D8 model. 
This is how vector meson dominance is realized in this model. 
The other constant $g_{a^n}$ is the coupling between an axial-vector meson 
$\tilde a_\mu^n$ and an external axial $U(1)$ field ${\cal A}_\mu$. 
This interaction does not contribute to electromagnetic form factors. 

The part of the 4d interaction lagrangian ${\cal L}_j$ that describes the cubic interaction among vector and axial-vector mesons is
\beqa
 {\cal L}^{4d}_{int} &\,=\,&  \tr \Big\{  \left(\partial^\mu {\tilde v}^{\nu\, n} - \partial^\nu {\tilde v}^{\mu \, n} \right) 
\Big( g_{v^n v^\ell v^m} [{\tilde v}_\mu^\ell, {\tilde v}_\nu^m]  
+ g_{v^n a^\ell a^m} [{\tilde a}_\mu^\ell, {\tilde a}_\nu^m]   \Big) \cr 
&& + g_{v^\ell a^m a^n} \left(\partial^\mu {\tilde a}^{\nu\, n} - \partial^\nu {\tilde a}^{\mu \, n} \right) 
 \Big( [{\tilde v}_\mu^\ell, {\tilde a}_\nu^m] - [{\tilde v}_\nu^\ell, {\tilde a}_\mu^m] \Big)  \Big\}\,, \label{IntLagrangian}
\eeqa

\noindent where the three-meson coupling constants are given by 
\beqa
g_{v^n v^\ell v^m} &\,=\,& \kappa \,  \int d \tilde z \,K(\tilde z)^{-1/3} 
\psi_{2n-1}(\tilde z) \psi_{2\ell -1}(\tilde z) \psi_{2m-1}(\tilde z) \label{couplingvvv}
\, ,\\
g_{v^\ell a^m a^n} &\,=\,& \kappa  \int d \tilde z \, K(\tilde z)^{-1/3} 
\psi_{2\ell -1}(\tilde z) \psi_{2m}(\tilde z)\psi_{2n}(\tilde z)  \, .\label{couplingvaa}
\eeqa

\noindent In order to obtain the masses $\,M_{v^n}\,$, $\,M_{a^n}\,$ and the 
coupling constants $\,g_{v^n}\,$, $\,g_{v^n v^\ell v^m}\,$ and $\,g_{v^\ell a^m a^n}\,$ 
we will now calculate numerically the wave functions $\psi_{r}(\tilde z)$. 

\subsection{Wave functions and Regge trajectories}

\FIGURE{
\epsfig{file=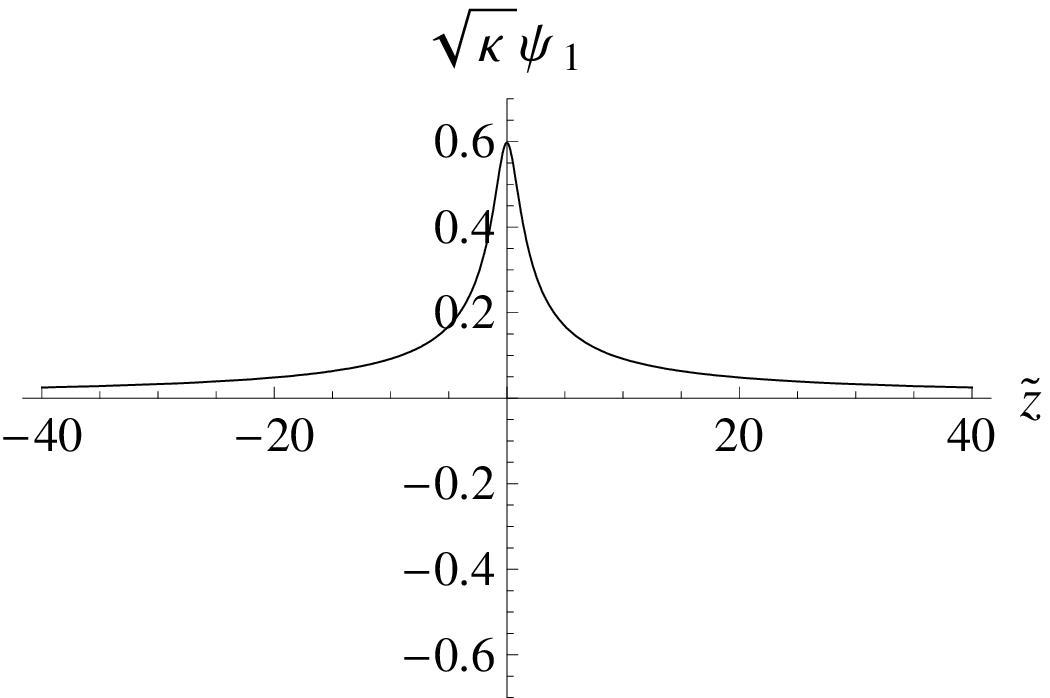,
width=5.5cm}
\epsfig{file=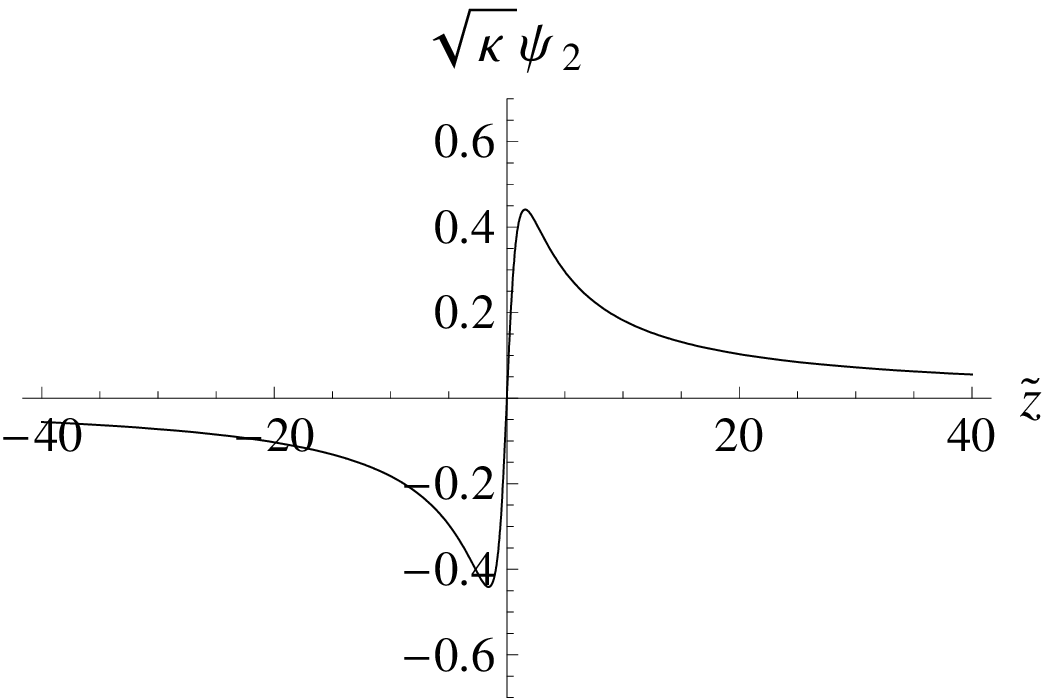,
width=5.5cm}
\epsfig{file=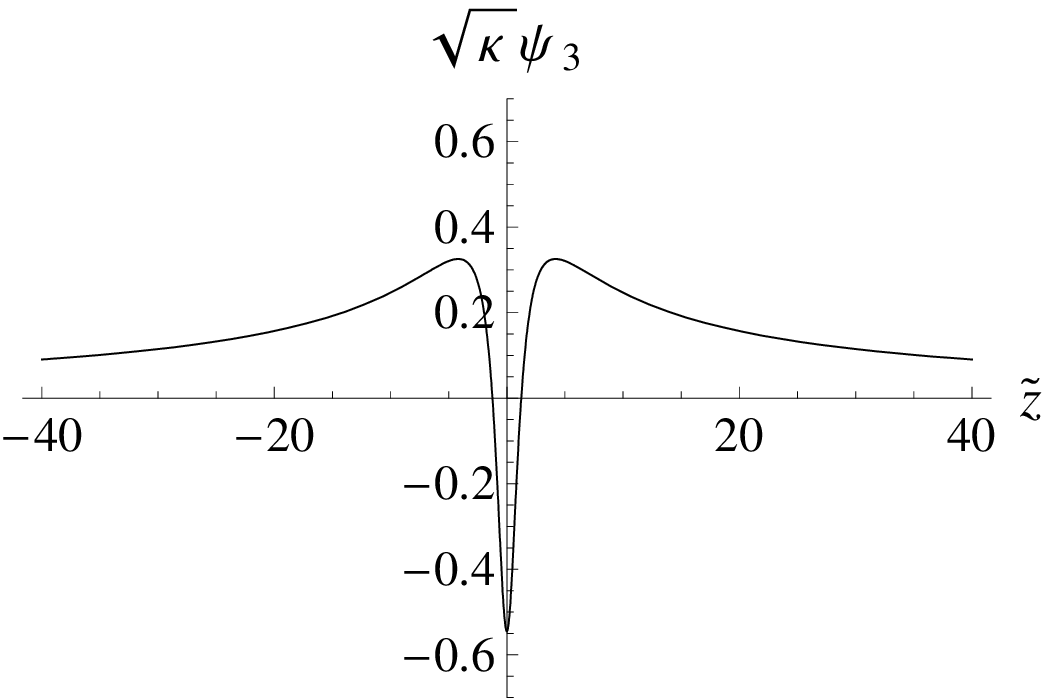,
width=5.5cm}
\epsfig{file=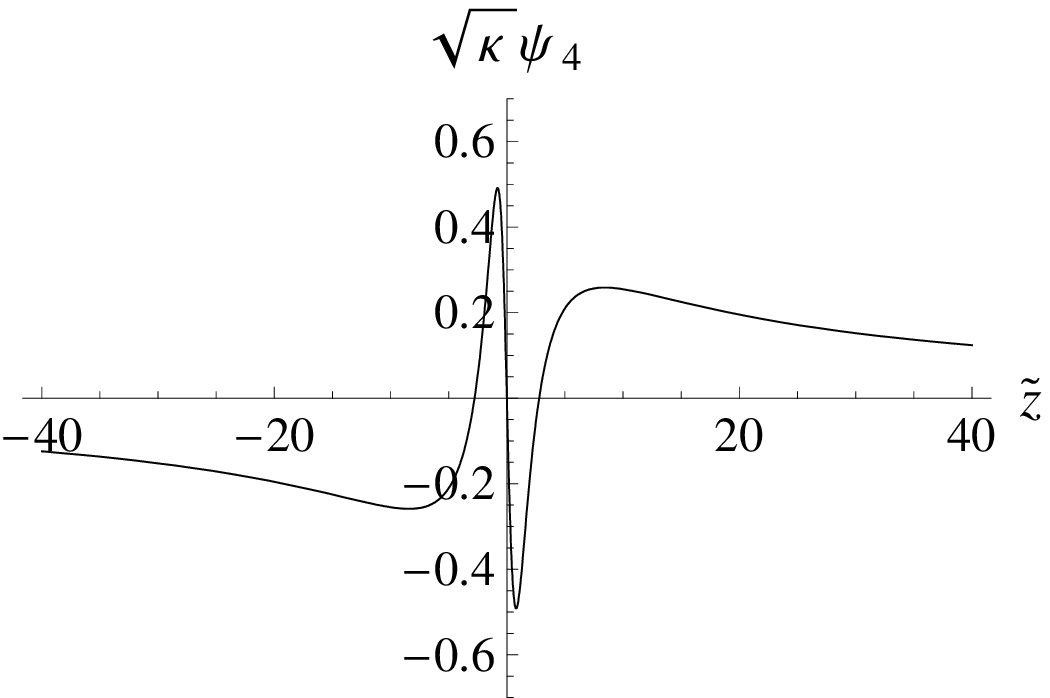,
width=5.5cm}
\epsfig{file=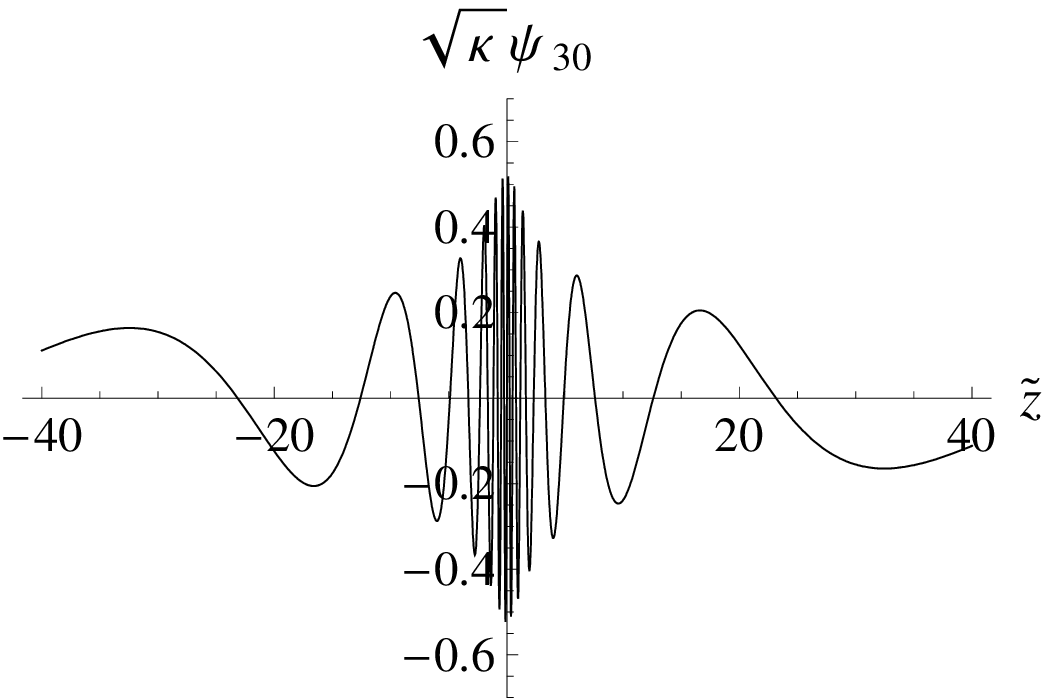,
width=5.5cm} 
\caption{Wave functions $\psi_{r}(\tilde z)$ multiplied by $\sqrt{\kappa}$ for the cases $r=1,2,3,4$ and $r=30$.}
\label{wavefunctions}}

Now we solve numerically the equations of motion for the vector and axial-vector modes using  the {\it shooting method}, as done  previously in \cite{Sakai:2004cn}. First, we analyze the large $\tilde z$ behavior of the $\psi_r(\tilde z)$ functions. From the normalization condition (\ref{vectormesonnorm}) and equation of motion (\ref{vectormesoneq}) we conclude that the $\psi_r$ functions decrease asymptotically as $\tilde z^{-1}$ when $\tilde z \to \pm \infty$. Then, it is convenient to define $\tilde\psi_r \equiv \tilde z\psi_r$. In terms of this function the equation of motion takes the form 
\beq
\tilde z \partial_{\tilde z} \left[ \tilde z \partial_{\tilde z} \tilde\psi_r \right] \,+\, A(\tilde z) \, \tilde z \partial_{\tilde z} \tilde\psi_r \,+\, B(\tilde z) \, \tilde\psi_r \,=\, 0 \, , \label{eqvectormesontilde}
\eeq
where 
\beqa
A(\tilde z) &\,=\,& \,-\, \frac{1 + 3 \tilde z^{-2}}{1 + \tilde z^{-2}} \, \equiv \, \sum_{\ell=0}^{\infty} A_\ell \, \tilde z^{-2 \ell /3} \, , \nonumber \\
B(\tilde z) &\,=\,& 2 \frac{\tilde z^{-2}}{1 + \tilde z^{-2}} \,+\, \lambda_r \tilde z^{-2/3} (1 + \tilde z^{-2} )^{-4/3} \, \equiv \, \sum_{\ell=0}^{\infty} B_\ell \, \tilde z^{-2 \ell /3} \,. \label{coeffexpansions}
\eeqa
The expansions defined in (\ref{coeffexpansions}) are well behaved for large $\tilde z$. They suggest the following ansatz for $\tilde\psi_r$ : 
\beq
\tilde\psi_r (\tilde z) \,=\, \sum_{\ell=0}^\infty \alpha_\ell \, \tilde z^{-2 \ell /3} \,, \label{ansatzsolution}
\eeq
with $\alpha_0=1$. Substituting (\ref{ansatzsolution}) in (\ref{eqvectormesontilde}) leads to the recurrence relations 
\beqa
\alpha_1 &\,=\,& \,-\, \frac{9}{10} \,  B_1 \, , \nonumber \\
\alpha_\ell &\,=\,& \left( \frac49 \ell^2 + \frac23 \ell \right)^{-1} \, \left( \frac23 \sum_{k=1}^{\ell-1} k \alpha_k A_{\ell-k} - \sum_{k=0}^{\ell-1} \alpha_k B_{\ell-k} \right) 
\eeqa

\FIGURE{
\epsfig{file=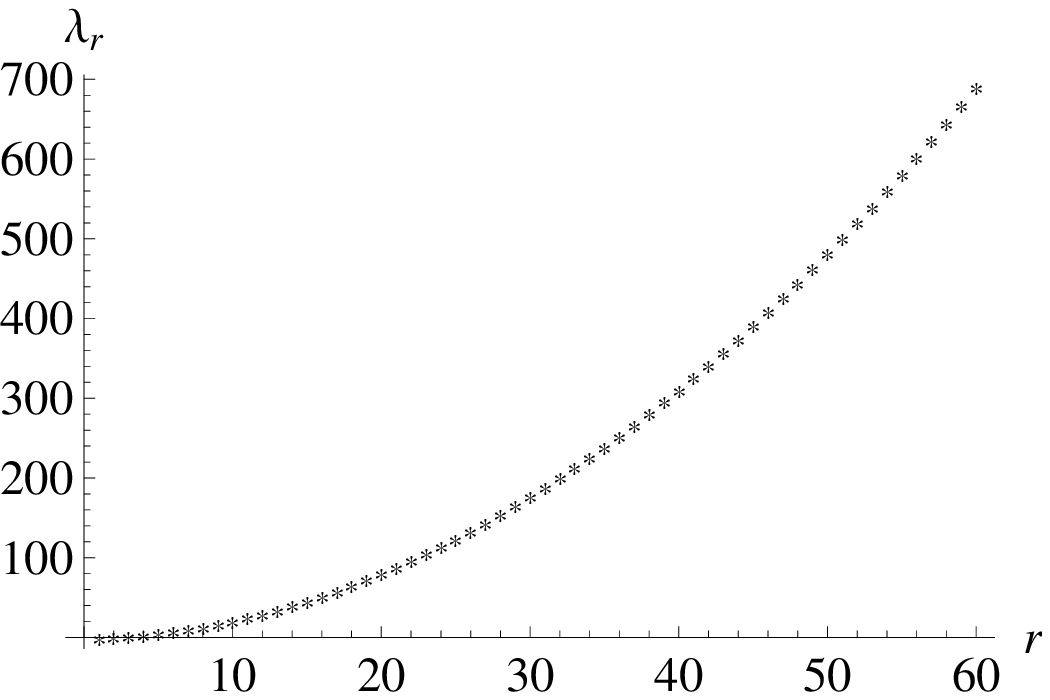,
width=6.7cm}
\epsfig{file=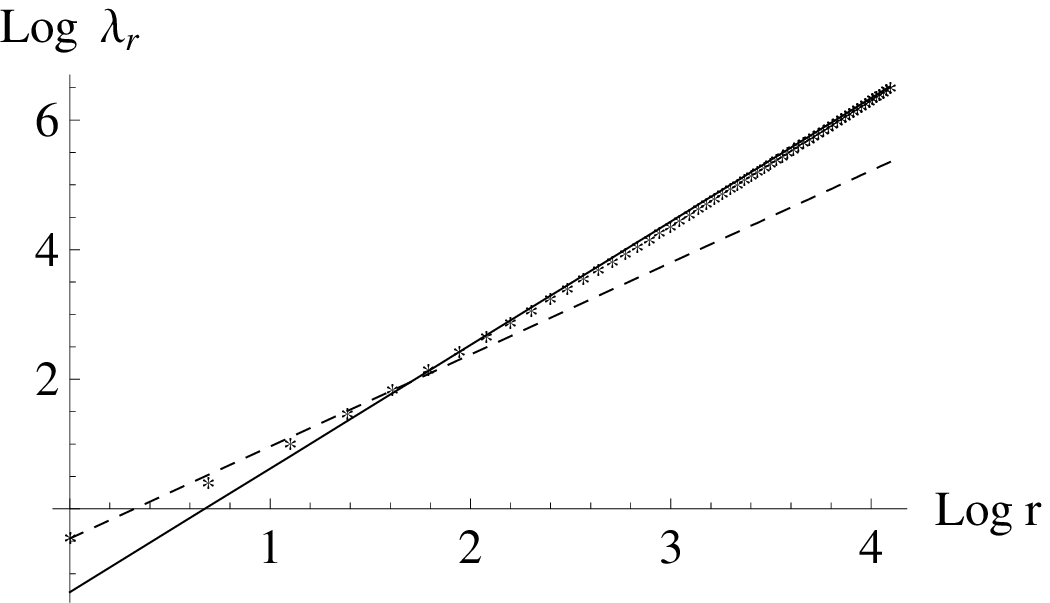,
width=7cm}
\caption{Vector and axial-vector meson Regge trajectory for D4-D8 model. The left panel shows the dependence of $\lambda_r$ with the radial number $r=1,2,3,..., 60$. In the right panel we plot a logarithmic graph of $\lambda_r$ and two linear fits in different ranges: the dashed line corresponds to $ -0.46 + 1.42 \, {\rm log}\, r$ for $r=\{1,5\}$,  while the solid line is $ -1.28 + 1.91 \, {\rm log}\, r$ for  $r=\{6,60\}$.}
\label{Regge}}

Now we analize the parity of the wave functions. 
The vector and axial-vector mesons transforms under parity in the following way 
\beq
v_\mu^n(t,-\overline x)\,=\,v_\mu^n(t,\overline x) \quad , \quad a_\mu^n(t,-\overline x)\,=\, -a_\mu^n(t,\overline x) \, .
\eeq
Such transformation exchange the gauge field chiral components, $A_L$ and $A_R$. In order to guarantee a a five dimensional gauge field invariance under the parity transformation $(t,\overline x, \tilde z) \to (t,-\overline x, -\tilde z)$ we impose the conditions 
\beq
\psi_{2n}(-\tilde z)\,=\, \psi_{2n}(\tilde z) \quad , \quad \psi_{2n-1}(-\tilde z)\,=\, - \psi_{2n-1}(\tilde z) \label{zparity}
\eeq
If the $\psi_{2n}(\tilde z)$ and $\psi_{2n-1}(\tilde z)$ wave functions are regular at the origin $\tilde z=0$ eq.  (\ref{zparity}) leads to the conditions 
\beq
\partial_{\tilde z} \psi_{2n}(0) \,=\, 0 \quad , \quad \psi_{2n-1}(0)\,=\, 0 \,. 
\eeq

Using these conditions at $\tilde z=0$ and the large $\tilde z$ behavior given by eq. (\ref{ansatzsolution}), 
we solved numerically eq. (\ref{vectormesoneq}), finding the solutions $\psi_r(z)$ and their eigenvalues $\lambda_r$ for $r=1$ to 60.
We plot some wavefunctions in Figure \ref{wavefunctions}. 
Note that the eigenfunctions $\psi_r(z)$ have an oscillatory behavior and 
the number of oscillations increases with $\lambda_r$. 
Looking at the small $z$ limit of equation (\ref{vectormesoneq}) one sees that it reduces to a sinusoidal wave equation with solutions:  
\beq
 \psi_{2n} \sim \sin \, (\sqrt{\lambda_{2n}\,} \, z) \,, \quad \quad \psi_{2n-1} \sim \cos\, (\sqrt{\lambda_{2n-1}\,} \, z) \,.  
\eeq

\noindent So, one expects an oscillatory behavior in this region. For large $z$ we observe the expected $z^{-1}$ behavior.

The eigenvalues $\lambda_r$ give us the vector and axial-vector quadratic masses $M_r^2= \lambda_r M_{\ss KK}^2$, according to equation (\ref{masses}). Notice that even values of $r$ correspond to axial-vector mesons and odd values of $r$ to vector mesons. 
In Figure \ref{Regge} we show the dependence of $\lambda_r$ on the radial number $r=1,2,3,..., 60$ . 
From the plots in this figure we conclude that vector and axial-vector mesons have the same radial Regge trajectory.

In the right panel of Figure \ref{Regge} we plot a logarithmic graph and present two linear fits for different ranges of $r$. 
For $r=1,..., 5$ the best linar fit is given by $ -0.46 + 1.42 \, {\rm log}\, r$, so that in this region $\lambda_r \sim r^{1.42}$.  
For the region $r=\{6,60\}$ the best fit is $ -1.28 + 1.91 \, {\rm log}\, r$ corresponding to $\lambda_r \sim r^{1.91}$. 
Fitting the data for larger values of $r$, we conclude that the Regge trajectory for vector and axial-vector mesons in the 
D4-D8 model approaches an asymptotic quadratic behaviour $\lambda_r \sim r^{2}$. 
Such behaviour is similar to the ones found in hard-wall model \cite{BoschiFilho:2002vd,BoschiFilho:2002ta,deTeramond:2005su,Erlich:2005qh,BoschiFilho:2005yh} 
and the D3-D7 brane model \cite{Kruczenski:2003be,Kirsch:2006he}.

\subsection{Coupling and decay constants}

\begin{table}
\bigskip\centerline{\begin{array}[b]{|c|c|c|c|c|c|}
\hline
 &&&&& \\
n&\frac{M_{v^n}^2}{M_{KK}^2}&\frac{g_{v^n}}{\sqrt{\kappa}M_{KK}^2}&\sqrt{\kappa}g_{v^nv^1v^1}& \sqrt{\kappa}g_{v^nv^2v^2}& \sqrt{\kappa}g_{v^nv^1v^2} \\[0.5ex]
&&&&& \\ 
\hline\hline
1 & 0.6693 	& 2.109  & 0.4466 & 0.2687  & -0.1465   \\
2 & 2.874 	& 9.108  & -0.1465 &  0.04261  & 0.2687  \\
3 & 6.591 	& 20.80  & 0.01843 &  0.1209 & -0.1477    \\
4 & 11.80 	& 37.15  & -3.689{\ss\times} 10^{\ss -4}& -0.1483 & 0.02371  \\
5 & 18.49 	& 58.17  & 2.695{\ss\times} 10^{\ss -4}&  0.03169 &  - 1.921 {\ss\times}10^{\ss -4}\\
6 & 26.67 	& 83.83  & 3.078{\ss\times} 10^{\ss -5}&  3.000 {\ss\times} 10^{\ss -4}& 4.469 {\ss\times}10^{\ss -4}     \\
7 & 36.34 	& 114.2  & 1.857{\ss\times} 10^{\ss-5}& 7.924 {\ss\times} 10^{\ss -4} &  7.560 {\ss\times}10^{\ss -5} \\
8 & 47.49  	& 149.1  & 6.996{\ss\times} 10^{\ss -6}& 1.824 {\ss\times} 10^{\ss -4}  & 4.080 {\ss\times}10^{\ss -5} \\
9 & 60.14 	& 188.7  & 3.508{\ss\times} 10^{\ss -6}& 9.417 {\ss\times} 10^{\ss -5}     & 1.723 {\ss\times}10^{\ss -5}     \\ \hline
\end{array}}\caption{Dimensionless squared masses and coupling constants for vector mesons.}
\label{vectortableconstants}
\end{table}

In order to calculate the elastic form factors for the lowest energy vector and axial-vector meson states, 
we must compute the vector meson decay constants $g_{v^n}$ and the coupling constants between the intermediate vector meson and the 
external vector mesons  $g_{v^nv^\ell v^m}$ and axial-vector mesons  $g_{v^n a^\ell a^m}$. 
We calculated these coupling constants using eqs. (\ref{decayvn}), (\ref{couplingvvv}), (\ref{couplingvaa})  
and our numerical results for the normalized wave functions $\psi_{2n}$ and $\psi_{2n-1}$ and the corresponding eigenvalues, 
for $n=1, 2, ..., 30$.

\begin{table}
\bigskip\centerline{\begin{array}[b]{|c|c|c|c|c|}
\hline
 & & & & \\
n & \frac{M_{a^n}^2}{M_{KK}^2} & \sqrt{\kappa}g_{v^na^1a^1} & \sqrt{\kappa}g_{v^na^2a^2} & \sqrt{\kappa}g_{v^na^1a^2}\\[0.5ex]
&&&& \\ 
\hline\hline
1& 1.569 &   0.2865 & 0.2572 & -0.1453\\
2& 4.546  &   0.1475 & 0.03770 &  0.1345   \\
3& 9.008 &  -0.1438 & 0.03294  &  0.1248   \\
4& 14.96 &   0.02617   & 0.1007 & -0.1465   \\
5& 22.39  &   8.738 {\ss\times}10^{\ss -5}   &  -0.1491  & 0.03023   \\
6& 31.32 &   5.312 {\ss\times}10^{\ss -4}  &  0.03552    &  1.995  {\ss\times}10^{\ss -4}  \\
7& 41.73 &   9.789 {\ss\times}10^{\ss -5}  &  6.837 {\ss\times}10^{\ss -4}& 7.289  {\ss\times}10^{\ss -4}      \\
8& 53.63 &   5.144{\ss\times}10^{\ss -5}  &  1.033 {\ss\times}10^{\ss -3} & 1.622 {\ss\times}10^{\ss -4}  \\
9& 67.02 &   2.212{\ss\times}10^{\ss -5}  &  2.721 {\ss\times}10^{\ss -4} & 8.421 {\ss\times}10^{\ss -5}  \\ \hline
\end{array}}\caption{Dimensionless squared masses and coupling constants for axial-vector mesons.}
\label{axialtableconstants}
\end{table}

We show some numerical results for vector mesons in Table \ref{vectortableconstants} and for axial-vector mesons in 
Table  \ref{axialtableconstants}. Note that the decay constants are positive, increase monotonically 
with $n$ and their ratio with the squared masses $g_{v^n}/M_{v^n}^2$  is approximately equal to $3.14 \sqrt{\kappa}$. 
On the other hand, the vector meson couplings $g_{v^nv^\ell v^m}$ and $g_{v^n a^\ell a^m}$ do not present a simple 
dependence with $n$ but oscillate (irregularly) with it. 
Nonetheless, these coupling constants obey sum rules that are crucial for the hadronic form factors, 
as we will discuss in the following section.


\section{Form Factors}
\label{Formfactors}

The eletromagnetic form factor represents the interaction of a particle with an external photon. 
Here we will calculate these form factors for vector and axial-vector mesons. 
As shown in \cite{Sakai:2005yt}, photon-meson-meson couplings are all cancelled in the D4-D8 model. 
This means that the photon interacts with a meson only through intermediate vector mesons. 
This is a realization of vector meson dominance (VMD) in electromagnetic scattering in this model.

In order to calculate the form factors we first obtain a general expression valid for the elastic 
and non-elastic cases. Then, for the elastic case, we calculate the electric, magnetic, 
quadrupole form factors as well as the longitudinal, transverse and 
longitudinal-transverse form factors. Also, for the first vector and axial-vector excitations,  
$\rho(770)$ and $a_1(1260)$, we calculate the eletric radius and the magnetic and quadrupole moments. 
We end this section with a brief discussion of the so called transition (non-elastic) form factors.



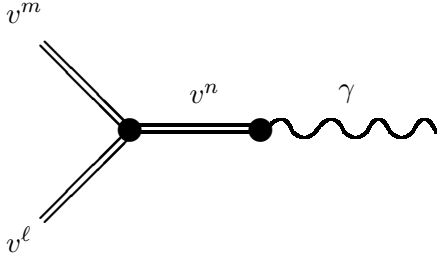
\begin{figure}
\begin{center}
\vskip 3.cm
\begin{picture}(0,0)(5,0)
\setlength{\unitlength}{0.06in}
\rm
\thicklines 
\put(-12,17){$v^m$}
\put(-9,15){\line(1,-1){7.5}}
\put(-8.6,15.4){\line(1,-1){7.5}} 
\put(-12,-3){$v^\ell$}
\put(-9,0){\line(1,1){7.5}}
\put(-8.6,-0.4){\line(1,1){7.5}}
\put(10.2,7.7){\circle*{2}}
\put(-1.2,8.1){\line(1,0){10.5}}
\put(-1.2,7.5){\line(1,0){10.5}}
\put(-1.2,7.7){\circle*{2}}
\put(4,10){$v^n$}
\put(17,10.5){$\gamma$}
\bezier{300}(10.5,7.5)(12.2,9.7)(13,7.5)
\bezier{300}(13,7.5)(14,6.5)(15,7.5)
\bezier{300}(15,7.5)(16.5,9.7)(17.5,7.5)
\bezier{300}(17.5,7.5)(18.5,6.5)(19.5,7.5)
\bezier{300}(19.5,7.5)(20.5,9.7)(21.5,7.5)
\bezier{300}(21.5,7.5)(22.5,6.5)(23.5,7.5)
\bezier{300}(23.5,7.5)(24.5,9.7)(25.5,7.5)
\end{picture}
\vskip 1.cm
\parbox{4.1 in}{\caption{Feynman diagram for vector meson form factor. A similar diagram holds for the axial-vector meson replacing the external lines $v^m, v^\ell$ by $a^m, a^\ell$.}}
\end{center}\label{Feynman}
\end{figure}
\vskip .5cm

The form factors are calculated from the matrix elements of the eletromagnetic current. 
The interaction of a vector meson with an off-shell photon is described by the matrix element
\begin{eqnarray}
\langle v^{m\,a}(p), \epsilon \vert {\tilde J}^{\mu c}(q) \vert v^{\ell \,b}(p'), \epsilon' \rangle 
&=&  (2\pi)^4 \delta^4(p'-p-q) \, \langle v^{m\,a}(p), \epsilon \vert J^{\mu c}(0) \vert v^{\ell \,b}(p'), \epsilon' \rangle \,,
\end{eqnarray}
\noindent where $v^{m}$ and  $v^{\ell}$ are the initial and final vector meson states with momenta $p$ and $p'=p+q$ and polarizations $\epsilon$ and $\epsilon'$. The operator ${\tilde J}^\mu$ is the Fourier transform of the electromagnetic current ${J}^\mu (x)$. 
This matrix element is calculated from the corresponding Feynman diagram shown in Figure 3. From the effective lagrangian 
(\ref{fourdimensionalefflag}), together with the interaction terms  (\ref{IntLagrangian}), we find 
\begin{eqnarray}
\langle v^{m\,a}(p), \epsilon \vert J^{\mu c}(0) \vert v^{\ell \,b}(p'), \epsilon' \rangle 
&=& \epsilon^\nu {\epsilon'}^\rho f^{abc} 
\left[ \eta_{\sigma\nu}(q-p)_\rho + \eta_{\nu\rho}(2p+q)_\sigma -  \eta_{\rho\sigma}(p + 2q)_\nu \right] \cr \cr
&& \times 
\sum_{n=1}^\infty {g_{v^n}g_{v^mv^nv^\ell}} \left[ \frac{\eta^{\mu\sigma} + \frac{q^\mu q^\sigma}{M_{v^n}^2}}{q^2+M_{v^n}^2} \right] 
\end{eqnarray}

\noindent where $f^{abc}$ is the structure constant of $U(N_f)$ and $M_{v^n}$ is the mass of the vector meson ${v^n}$. 
Using the sum rule 
\begin{equation}
\sum_{n=1}^\infty \frac{g_{v^n}g_{v^nv^mv^\ell}}{M_{v^n}^2} = \delta_{m\ell}
\label{sumrule}
\end{equation}
 
 \noindent obtained in \cite{Sakai:2005yt}, we find  
 \begin{eqnarray}
\langle v^{m\,a}(p), \epsilon \vert J^{\mu c}(0) \vert v^{\ell \,b}(p'), \epsilon' \rangle 
&=& \epsilon^\nu {\epsilon'}^\rho f^{abc} 
\left[ \eta_{\sigma\nu}(q-p)_\rho + \eta_{\nu\rho}(2p+q)_\sigma -  \eta_{\rho\sigma}(p + 2q)_\nu \right] \cr \cr
&& \times \left\{ \left( {\eta^{\mu\sigma} - \frac{q^\mu q^\sigma}{q^2}} \right) F_{v^mv^\ell}(q^2) 
+ \delta_{m\ell} \frac{q^\mu q^\sigma}{q^2} \right\} 
\label{Formlong}
 \end{eqnarray}
 
 \noindent where the generalized vector meson form factor is defined by 
\begin{equation}
F_{v^mv^\ell}(q^2)=\sum_{n=1}^\infty
\frac{g_{v^n}g_{v^nv^mv^\ell}}{q^2+M_{v^n}^2} \ .
\label{form:vn}
\end{equation} 

Taking into account the transversality of the vector meson polarizations: 
$\epsilon \cdot p =0 = \epsilon' \cdot p' $, we obtain 
 \begin{eqnarray}
&& \langle v^{m\,a}(p), \epsilon \vert J^{\mu c}(0) \vert v^{\ell \,b}(p'), \epsilon' \rangle \cr\cr
&& \qquad = \epsilon^\nu {\epsilon'}^\rho f^{abc} 
\left[ \eta_{\nu\rho}(2p+q)_\sigma  + 2(\eta_{\sigma\nu} q_\rho - \eta_{\rho\sigma}q_\nu) \right] 
 \left( {\eta^{\mu\sigma} - \frac{q^\mu q^\sigma}{q^2}} \right) F_{v^mv^\ell}(q^2) \,.
\label{Formlong2}
 \end{eqnarray}
 
\noindent Note that the term  involving the factor $\delta_{m\ell}$ in eq. (\ref{Formlong}) 
did not contribute since in the elastic case ($m=\ell$) we have: $2p\cdot q + q^2 =0$.  Note also 
that this  matrix element satisfies the tranversality condition 
$q_\mu \langle v^{m} \vert J^{\mu}(0) \vert v^{\ell} \rangle =0$. 

In a similar way, for axial-vector mesons we can calculate the form factors from the matrix element
$\langle a^{m\,a}(p), \epsilon \vert J^{\mu c}(0) \vert a^{\ell \,b}(p'), \epsilon' \rangle$. 
This corresponds to evaluating Feynman diagrams similar to Fig. 3, but with the external vector meson lines
replaced by the axial-vector mesons $a^m$ and $a^\ell$. Note that the internal vector meson line $v^n$, 
representing vector meson dominance, is unchanged. Thus, the generalized axial-vector meson form factor is
\begin{equation}
F_{a^ma^\ell}(q^2)=\sum_{n=1}^\infty
\frac{g_{v^n}g_{v^na^ma^\ell}}{q^2+M_{v^n}^2} \ .
\label{form:an}
\end{equation}

\subsection{Elastic case}

The elastic form factor for vector mesons can be obtained considering the previous calculation with the 
same vector meson $v^m$ in the initial and final states. Then, from eq. (\ref{Formlong2}) we find 
\begin{eqnarray}
&& \langle v^{m\,a}(p), \epsilon \vert J^{\mu c}(0) \vert v^{m \,b}(p'), \epsilon' \rangle \cr\cr
&& \qquad = f^{abc} \left\{ (\epsilon \cdot \epsilon') (2p+q)^\mu  + 
2\left[ \epsilon^\mu (\epsilon'\cdot q) - {\epsilon'}^\mu (\epsilon \cdot q) \right] \right\}
F_{v^m}(q^2) \,, 
\label{Formelastic}
 \end{eqnarray}

\noindent where $F_{v^m}(q^2)$ is the elastic form factor:

\begin{equation}
F_{v^m}(q^2)=\sum_{n=1}^\infty
\frac{g_{v^n}g_{v^nv^mv^m}}{q^2+M_{v^n}^2} \ .
\label{form:vnelast}
\end{equation} 

\noindent Note that eqs. (\ref{Formelastic}) and (\ref{form:vnelast}) are also valid for 
the axial-vector mesons, replacing $v^m$ by $a^m$. We calculated numerically these sums from $n=1$ to
$n=30$ using the results obtained for the masses and couplings. We adopted $M_{\ss KK}= 0.946$ GeV, 
as in ref. \cite{Sakai:2004cn}. 
We plot in Figure \ref{v1a1formfactors} the elastic form factors for the vector meson 
$\rho$(770) ($v^1$) and axial-vector meson $a_1$(1260) ($a^1$). Note that when $q^2\to 0$, 
the vector and axial-vector form factors go to one, thanks to the sum rule (\ref{sumrule}).

\FIGURE{
\epsfig{file=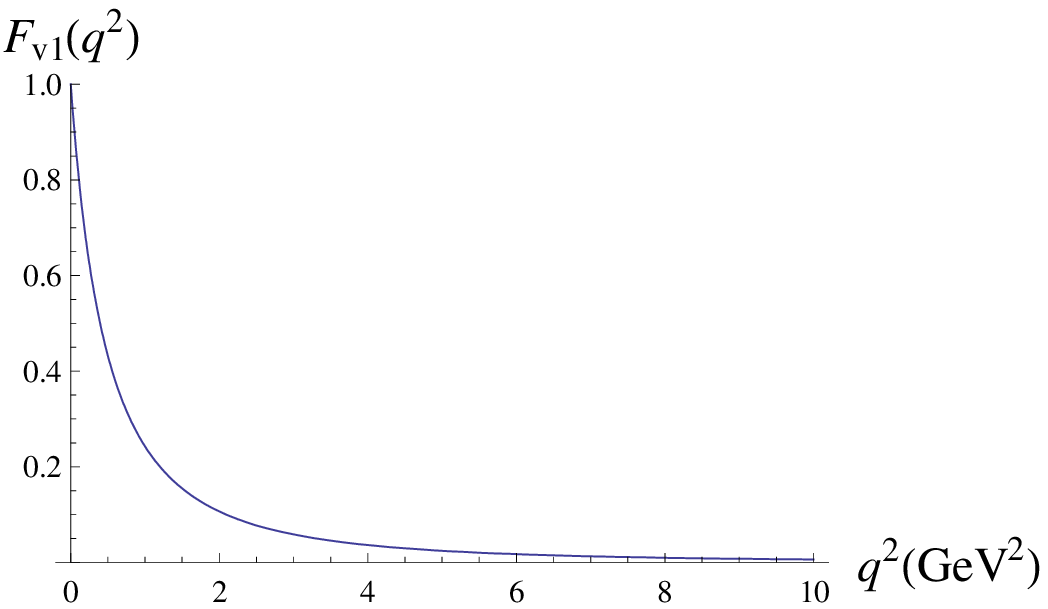, width=7cm} 
\epsfig{file=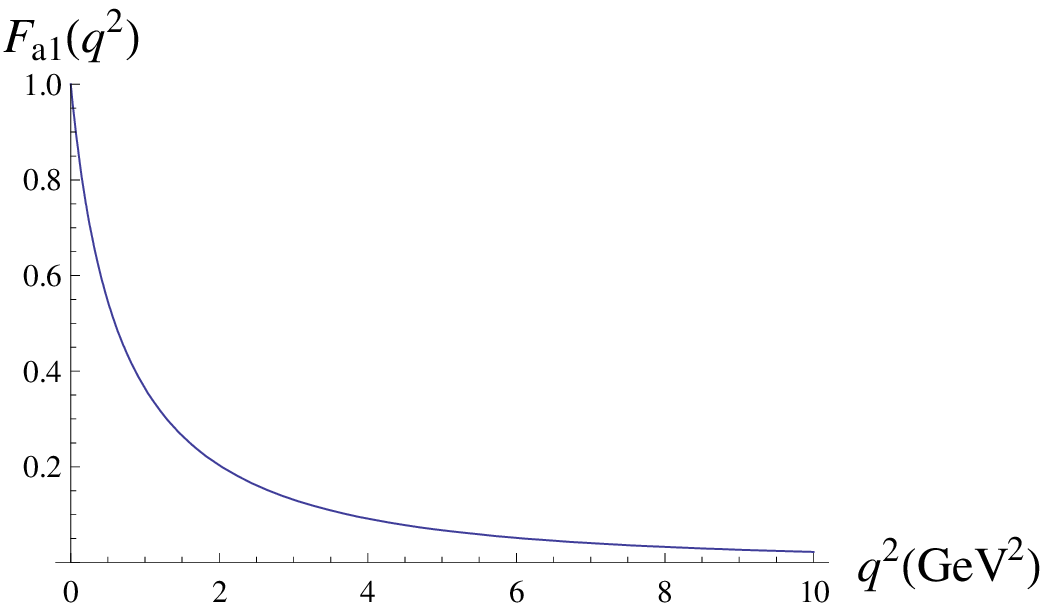, width=7cm} 
\caption{Elastic form factors for $v^1$ and $a^1$.} 
\label{v1a1formfactors}}

\FIGURE{
\epsfig{file=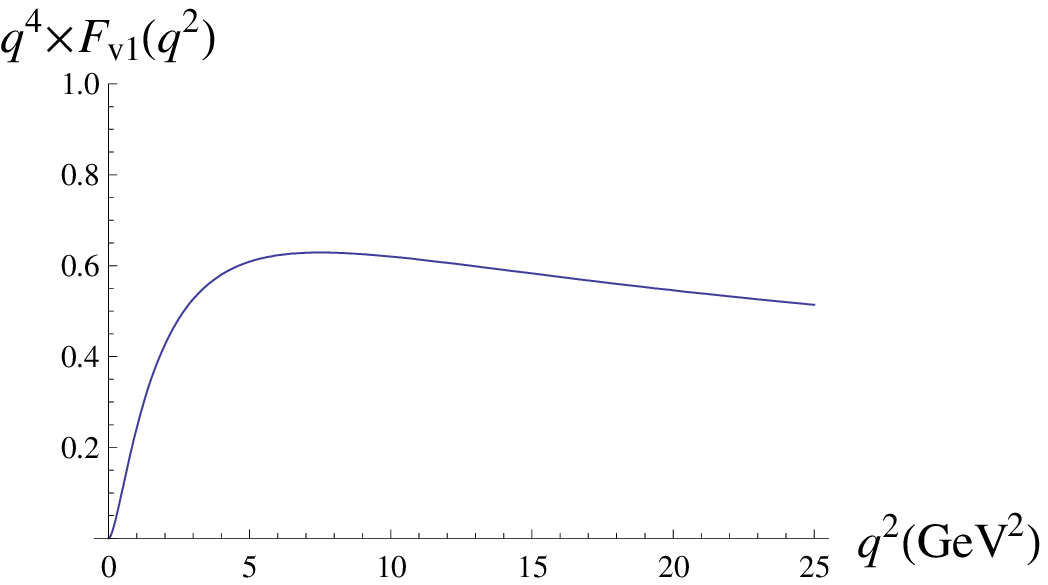, width=7cm} 
\epsfig{file=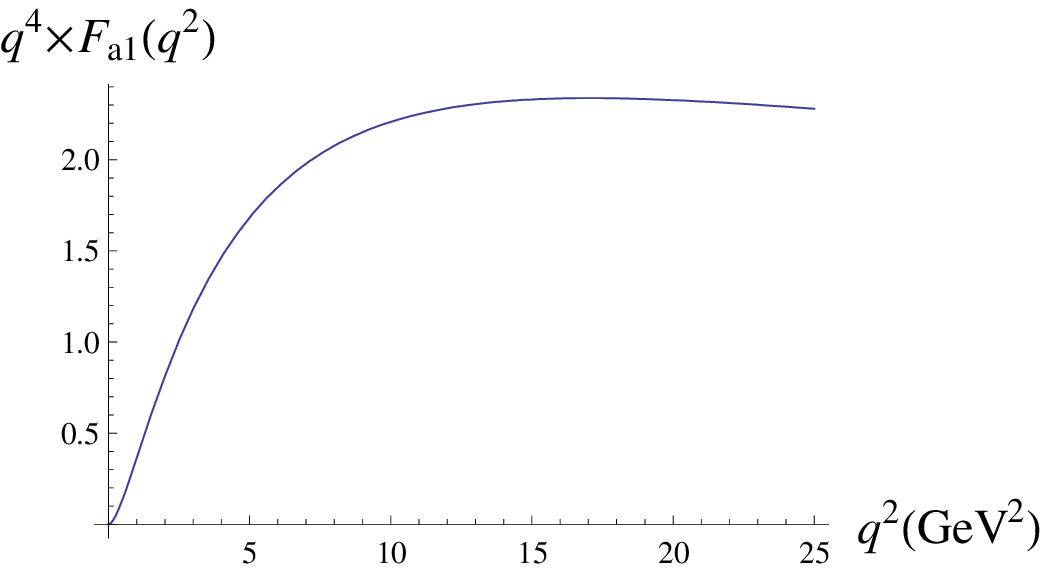, width=7cm} 
\caption{$q^4$ times the elastic form factors for $v^1$ and $a^1$.} 
\label{Q4formfactors}}

Let us now investigate the large $q^2$ behaviour of the elastic factors.
Performing an expansion in powers of $q^{-2}$:  
\begin{eqnarray}
F_{v^m}(q^2) &=& \frac{1}{q^{2}}\sum_{n=1}^\infty g_{v^n}g_{v^nv^mv^m} \left( 1 - \frac{M_{v^n}^2}{q^2} + {\cal O}(q^{-4}) \right) \nonumber\\
F_{a^m}(q^2) &=& \frac{1}{q^{2}}\sum_{n=1}^\infty g_{v^n}g_{v^na^ma^m} \left( 1 - \frac{M_{v^n}^2}{q^2} + {\cal O}(q^{-4}) \right)  
\ ,  \label{expelast}
\end{eqnarray} 

\noindent we see that the dominant terms would be of order $q^{-2}$. We calculated the coefficients of these terms  
using our numerical results with $n=1,...,30$. We found
 \begin{eqnarray}
\sum_{n=1}^{30}
{g_{v^n}g_{v^nv^1v^1}} &\approx&  0.0005835({\rm GeV})^2
 \quad ,\quad \sum_{n=1}^{30}
{g_{v^n}g_{v^na^1a^1}} \approx 0.0001485({\rm GeV})^2\,,
\cr
\sum_{n=1}^{30}
{g_{v^n}g_{v^nv^2v^2}} &\approx& -0.0002939({\rm GeV})^2
\quad ,\quad
\sum_{n=1}^{30}
{g_{v^n}g_{v^na^2a^2}} \approx  -0.002043({\rm GeV})^2
\cr
\sum_{n=1}^{30}
{g_{v^n}g_{v^nv^3v^3}} &\approx & -0.003339({\rm GeV})^2
\quad ,\quad 
\sum_{n=1}^{30}
{g_{v^n}g_{v^na^3a^3}} \approx -0.007522({\rm GeV})^2\,.
\label{Superconv123}
\end{eqnarray}

\noindent These results indicate that the following superconvergence relations hold in the D4-D8 model: 
\begin{equation}
\sum_{n=1}^\infty
{g_{v^n}g_{v^nv^mv^m}} = 0\; \; ,  \;\;
\sum_{n=1}^\infty
{g_{v^n}g_{v^na^ma^m}} = 0 \,,
\label{superconv}
\end{equation}

\noindent so, from eqs. (\ref{expelast}) we expect that the form factors decrease 
approximately as $q^{-4}$, for large $q^2$.  

In order to investigate this behaviour, we plot in Figure \ref{Q4formfactors} 
the form factors for the first vector and axial-vector states multiplied by $q^4$. 
We also plotted the elastic form factors for the first three excited states multiplied by 
$q^4 $ in Figure \ref{Q4Formfactorhigher}.

\FIGURE{
\epsfig{file=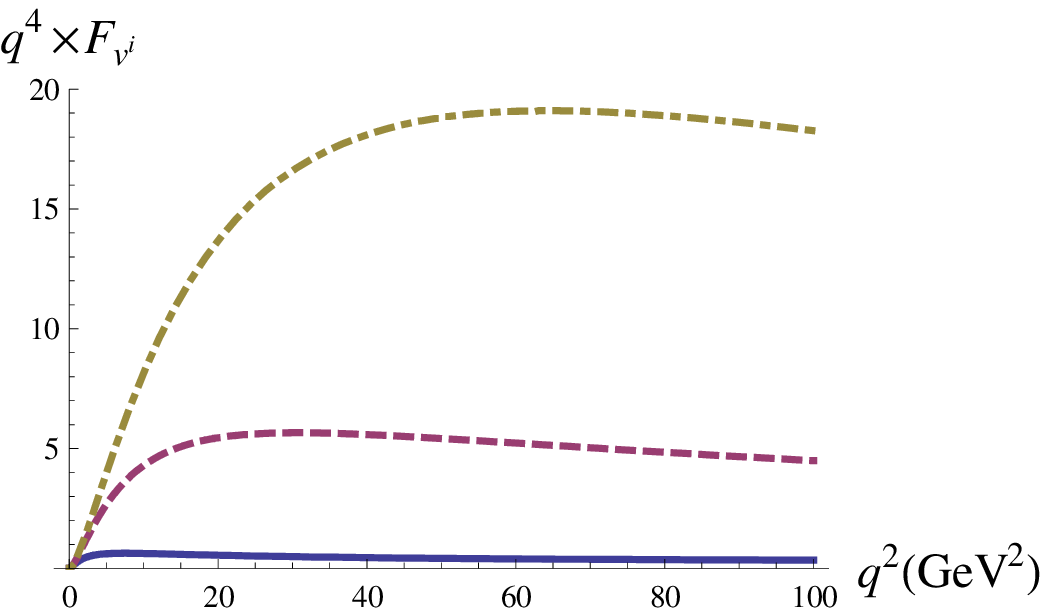, width=7cm} 
\epsfig{file=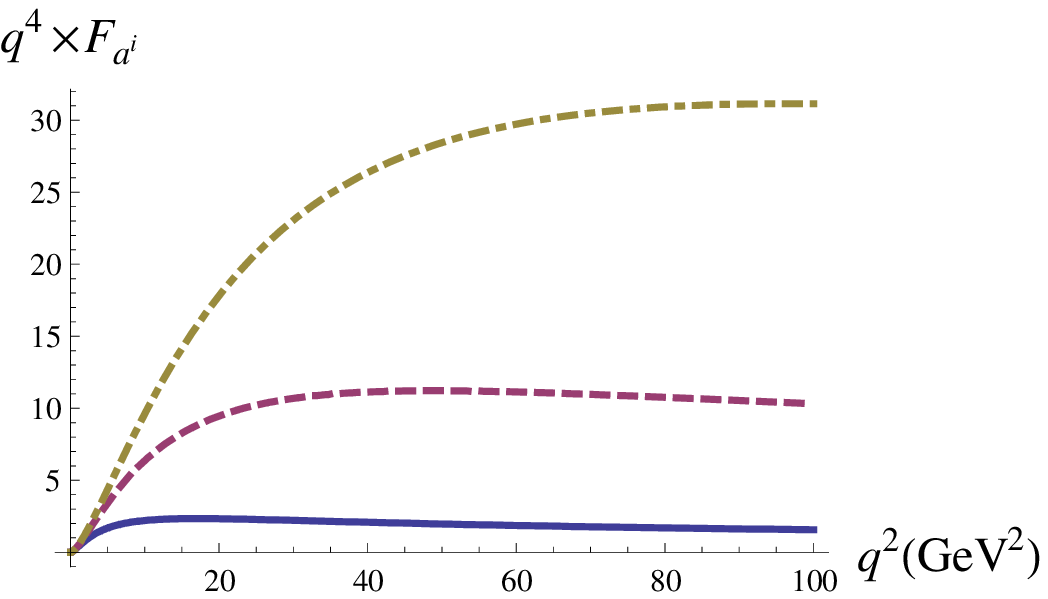, width=7cm} 
\caption{$q^4$ times the elastic form factors for vector and axial-vector mesons 
for the first three states: $i$ =1 (solid line), 2 (dashed line), 3 (dot-dashed line). 
Left panel: $ F_{v^i}$. Right panel $ F_{a^i}$.} 
\label{Q4Formfactorhigher}}

We conclude that the form factors approximately decrease as $q^{-4}$, for large $q^2$, 
within the numerical errors.  The deviation from the $q^{-4}$ dependence is associated 
with the non-vanishing of the numerical sums (\ref{Superconv123}).


\bigskip

\noindent {\bf Electric, magnetic and quadrupole form factors}

\smallskip

The matrix element of the electromagnetic current for a spin one particle in the elastic case  
can be decomposed as\cite{Grigoryan:2007vg}
\begin{eqnarray}
 \langle p, \epsilon \vert J_{EM}^{\mu }(0) \vert p', \epsilon' \rangle 
&=&   (\epsilon \cdot \epsilon') (2p+q)^\mu F_1 ( q^2 )  + 
\left[ \epsilon^\mu (\epsilon'\cdot q) - {\epsilon'}^\mu (\epsilon \cdot q) \right] 
\left[ F_1 (q^2) + F_2(q^2) \right]  \nonumber\\
&+&  \frac{1}{p^2} ( q\cdot \epsilon') (q \cdot \epsilon ) (2p+q)^\mu F_3 (q^2) \,
.
\label{FormDecomp}
 \end{eqnarray}
 
\noindent From $F_1 , F_2 $ and $F_3$ we can define
\begin{equation}
F_E = F_1 + \frac{q^2}{6 p^2} \Big[ F_2 - ( 1 - \frac{q^2}{4 p^2})  F_3 \Big]
\,\,\, , \,\,\,
F_M = F_1 + F_2\,,
\,\,\, , \,\,\, 
F_Q = - F_2 + \Big( 1 - \frac{q^2}{4p^2} \Big) F_3 
\end{equation}

\noindent where $F_E, F_M $ and $F_Q$ are the electric, magnetic, and quadrupole form factors.
From eqs. (\ref{Formelastic}) and (\ref{FormDecomp}) we find that for a vector meson $v^m$
\begin{equation}
F_1^{(v^m)} = F_2^{(v^m)} =  F_{v^m}\,\,,\,\,\;\;\;\; F_3^{(v^m)} = 0 \,,
\end{equation}

\noindent where $F_{v^m}$ is given by eq. (\ref{form:vnelast}). 
Hence the electric, magnetic and quadrupole form factors predicted by the D4-D8 brane model 
for vector mesons are
\begin{equation}
F_E^{(v^m)} = ( 1  + \frac{q^2}{6 p^2} ) F_{v^m}
\,\,\, , \,\,\,
F_M^{(v^m)} = 2 F_{v^m} \,,
\,\,\,  \,\,\,
F_Q^{(v^m)} = - F_{v^m}  \,.
\end{equation}

\noindent The same results hold for the axial-vector mesons $a^m$.
From these form factors we can estimate three important physical quantities associated with 
the vector mesons:  the electric radius, the magnetic and quadrupole moments.


The electric radius for the vector and axial-vector mesons are given by  
\begin{equation}
\langle r^2_{v^m}\rangle = -6\frac{\mathrm{d}}{\mathrm{d}q^2}F_E^{(v^m)}(q^2)|_{q^2=0}\;\; , \;\; 
\langle r^2_{a^m}\rangle = -6\frac{\mathrm{d}}{\mathrm{d}q^2}F_E^{(a^m)}(q^2)|_{q^2=0}\,.
\end{equation}

\noindent Using our numerical results for the form factors for the lowest excited states $\rho$ and $a_1$,
we find the electric radii:
\begin{equation}
 \langle r^2_{\rho}\rangle = 0.5739 \,{\rm fm}^2\;\;,\;\;\;\;\;\; \langle r^2_{a_1}\rangle = 0.4061 \,{\rm fm}^2 \,.
\end{equation}




The magnetic and quadrupole moments are defined by
\begin{equation}
\mu \, \equiv \, F_M(q^2)|_{q^2=0}\;\; , \;\;\;\;\;\; 
D \, \equiv \,  - \frac{1}{p^2} F_Q(q^2)|_{q^2=0}\;.
\end{equation}

Using the fact that $F_{v^m}$ and $F_{a^m}$ go to one when $q^2 \to 0$, we obtain
\begin{eqnarray}
\mu_{v^m} = \mu_{a^m} = 2  \quad , \quad 
D_{v^m} = - \frac{1}{M_{v^m}^2} \quad , \quad 
D_{a^m} = - \frac{1}{M_{a^m}^2} \; .
\end{eqnarray}

Our results for electric radius, magnetic and quadrupole moments for the vector meson  $\rho$ 
are in agreement with the hard wall model results found in \cite{Grigoryan:2007vg}.

\bigskip

\noindent {\bf Decomposition in terms of transverse and longitudinal polarizations} 
\smallskip

It is interesting to calculate also the form factors of vector mesons with specific polarizations
\cite{Ioffe:1982qb}. A vector particle with mass $M$ and four momentum  $ ( E, 0, 0, p_z ) $ has three 
independent polarizations that can be written as 
\begin{equation}
\epsilon^1_T = ( 0, 1, 0, 0 )\quad , \quad \epsilon^2_T = ( 0, 0, 1, 0 )\quad , \quad
\epsilon_L = ( p_z/M , 0, 0, E/M ) \,\, .
\end{equation}

When studying the elastic scattering of a photon and a vector meson in the Breit frame,
the initial and final four momenta of the vector meson take, respectively, the forms 
$p = ( E, 0, 0, -  q/2 )\,$, and $\,p' = ( E, 0, 0,  q/2 ) \,$. In this frame we define the form 
factors
\begin{eqnarray}
F_{TT} ( q^2 ) &=&   \frac{\langle p, \epsilon_T \vert J_{0} (0) \vert p', \epsilon'_T \rangle}{ 2E} 
\qquad, \qquad  
F_{LT} ( q^2 ) =  \frac{ \langle p, \epsilon_T \vert J_{x} (0) \vert p', \epsilon'_L \rangle }{ 2E}\cr\cr
F_{LL} ( q^2 ) &=&  \frac{ \langle p, \epsilon_L \vert J_{0}(0) \vert p', \epsilon'_L \rangle }{2E}\,.
\end{eqnarray}

\noindent From eq. (\ref{Formelastic}) we find 
\begin{equation} 
F_{TT}^{(v^m)} =  F_{v^m}\quad ,\quad F_{LT}^{(v^m)} = \frac{q}{M_{v^m}} F_{v^m}
\quad ,\quad F_{LL}^{(v^m)} = \left( 1 - \frac{q^2}{2 M^2_{v^m}} \right) F_{v^m}\,.
\end{equation}

The large  $q^2$ behavior of these form factors in the D4-D8 model is determined
from $ F_{v^m}$, eq. (\ref{form:vnelast}). 
We obtain the asymptotic behaviors: $ F_{TT}^{(v^m)} \sim q^{-4} \,,\,
F_{LT}^{(v^m)} \sim q^{-3} \,,\, F_{LL}^{(v^m)} \sim q^{-2} \,$, in agreement with QCD calculations 
(see for instance \cite{Ioffe:1982qb}).

\subsection{Non-elastic case}

When the initial and final meson states are different, the interaction with the photon is described by 
the generalized form factors of eqs.  (\ref{form:vn}), (\ref{form:an}), when $m\not=\ell$, 
also known as transition form factors. 
We calculated these form factors for an initial vector meson in the lowest state $v^1$ and final states 
$v^2, v^3, v^4 $. We also calculated the  transitions form factors for the initial axial-vector meson 
$a^1$ and final  $ a^2, a^3, a^4$.

We illustrate in Figure \ref{formfactorsinelastic}
the dependence of these form factors with the momentum transfer $q^2$, 
compared with the corresponding elastic form factors. 
We see that the transition form factors vanish as $q^2\to 0$, 
in contrast to the elastic form factors
that approach the unit in this limit.
Note that as $q^2$ increases from zero, first $F_{v^1}$ dominates, 
then $F_{v^1v^2}$ dominates, then $F_{v^1v^3}$, and so on.
The same situation occurs for the axial-vector case. 
Physically, this corresponds to the fact that as the momentum transfer increases,
the amplitude for producing heavier final states increases as well. 

\FIGURE{
\epsfig{file=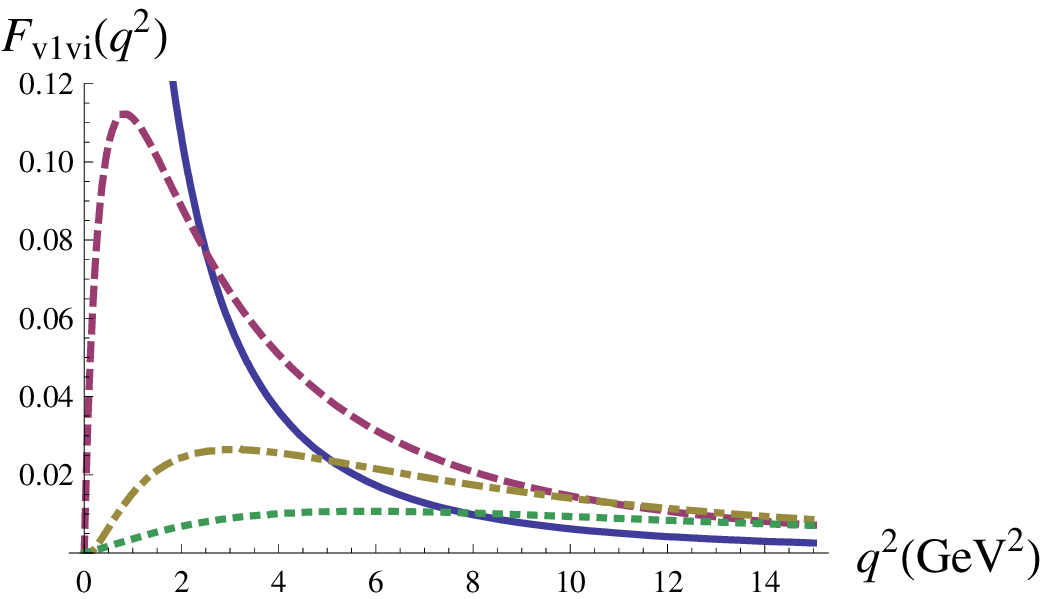, width=7cm} 
\epsfig{file=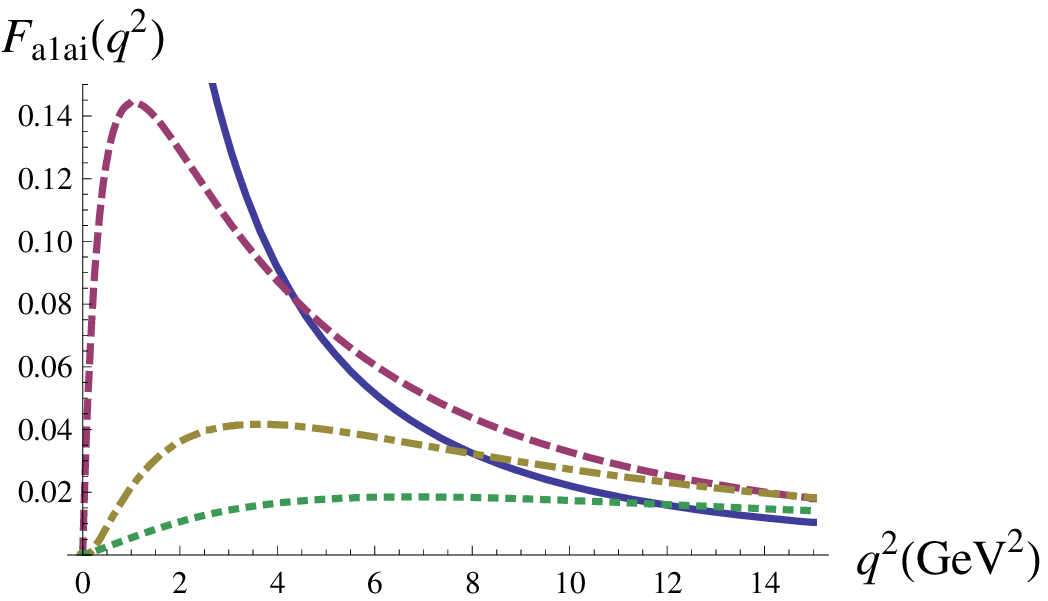, width=7cm} 
\caption{Transition form factors. Left panel:  $F_{v^1v^i}$, right panel: $F_{a^1a^i}$,  
for $i$ =1 (solid line), 2 (dashed line), 3 (dot-dashed line), 4 (dotted line). 
The solid lines corresponding to $i=1$ approaches one as $q^2\to 0$.} 
\label{formfactorsinelastic}}

\FIGURE{
\epsfig{file=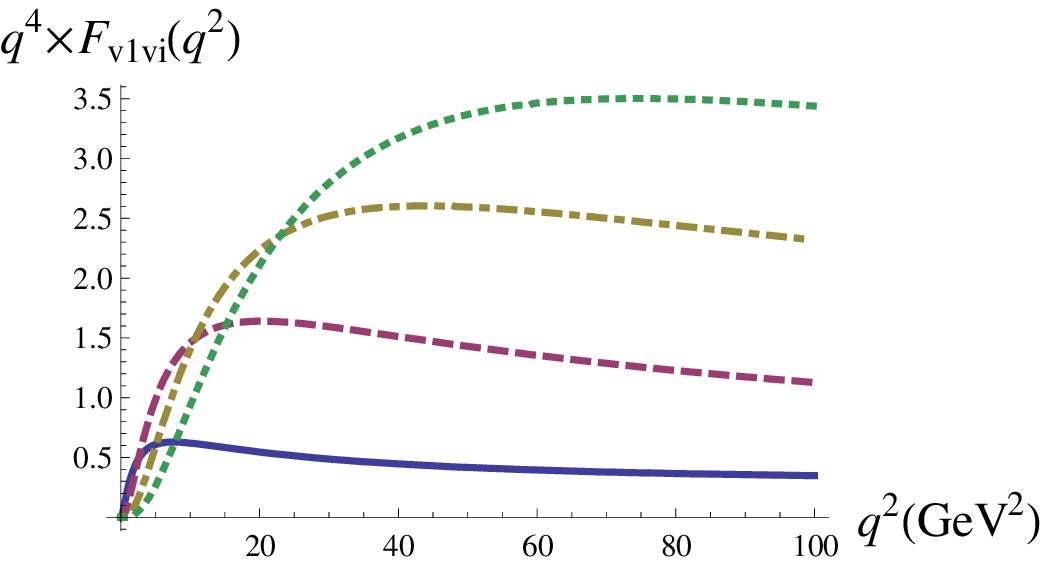, width=7cm} 
\epsfig{file=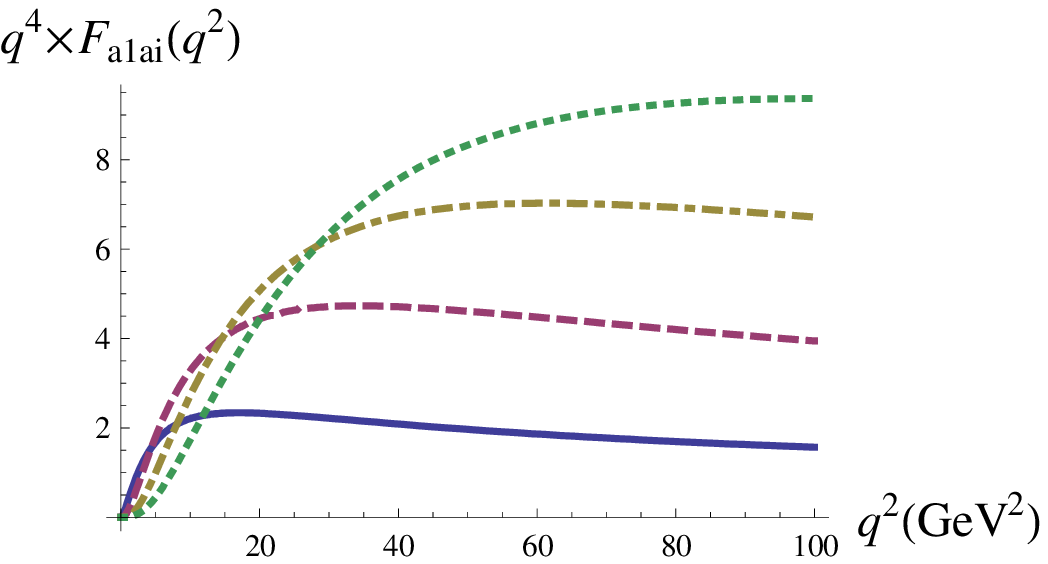, width=7cm} 
\caption{$q^4$ times the form factors. Left panel:  $F_{v^1v^i}$, right panel: $F_{a^1a^i}$,  
for $i$ =1 (solid line), 2 (dashed line), 3 (dot-dashed line), 4 (dotted line).} 
\label{Q4formfactorsinelastic}}

Regarding the large  $q^2 $ dependence of these non-elastic form factors,  
we can make an expansion similar to eq. (\ref{expelast}) but for different initial and final states. 
If the following superconvergence relations hold: 
 \begin{equation}
\sum_{n=1}^\infty
{g_{v^n}g_{v^nv^mv^l}} = 0 \; \; ,  \;\;
\sum_{n=1}^\infty
{g_{v^n}g_{v^na^ma^l}} = 0\,.
\label{superconvinelast}
\end{equation}

\noindent we should find a $ q^{-4} $ behavior as we found in the elastic case. 
We calculated these sums from $n=1,..., 30$, for the relevant states, finding:
\begin{eqnarray}
    \sum_{n=1}^{30}
    {g_{v^n}g_{v^nv^1v^2}} &\approx& -0.000367 ({\rm GeV})^2
,\quad 
    \sum_{n=1}^{30}
    {g_{v^n}g_{v^na^1a^2}} \approx -0.0007793 ({\rm GeV})^2
,\cr 
    \sum_{n=1}^{30}
    {g_{v^n}g_{v^nv^1v^3}} &\approx& -0.0002850 ({\rm GeV})^2
,\quad 
    \sum_{n=1}^{30}
    {g_{v^n}g_{v^na^1a^3}} \approx  -0.001863 ({\rm GeV})^2
,\cr 
    \sum_{n=1}^{30}
    {g_{v^n}g_{v^nv^1v^4}} &\approx& -0.001190 ({\rm GeV})^2
,\quad 
    \sum_{n=1}^{30}
    {g_{v^n}g_{v^na^1a^4}} \approx -0.001484 ({\rm GeV})^2
 \,.    
\end{eqnarray}

\noindent These results indicate the validity of relations (\ref{superconvinelast}).

In order to check the large $q^2$ behavior, we plot in Figure \ref{Q4formfactorsinelastic}
the transition form factors multiplied by $q^4$, and the corresponding elastic form factors for comparison. 
We conclude that the transition form factors approach asymptotically the expected $q^{-4}$ dependence.

\section{Conclusion}

In this work we investigated the electromagnetic scattering of 
vector and axial-vector mesons in the D4-D8 model. 
We calculated the elastic and transition meson form factors and found 
the expected behaviour at small and large momentum transfer $q^2$. 
In the elastic case, we calculated the magnetic and quadrupole moments 
and, for the mesons $\rho$ and $a_1$,  the electric radii. 
We also analized the momentum dependence of $ F_{TT}\,, F_{LT}\,$ and 
$\, F_{LL}\,$ at large $q^2$ finding a QCD like asymptotic behaviour. 

Our analysis of the form factor momentum dependence led us to non-trivial sum rules
for the vector and axial-vector coupling constants. These sum rules generalize the 
superconvenge rule previously obtained within the hard wall model for the elastic case  
\cite{Grigoryan:2007vg}.

Although the D4-D8 brane model was originally proposed to describe the low energy regime of QCD,
our results suggest that vector meson form factors at intermediate and high energies show a $q^2$ 
dependence consistent with QCD expectations. It would be interesting to investigate if other aspects
of hadronic scattering  at high energy could be described by this holographic model. 

\bigskip

\noindent {\bf Acknowledgments:} The authors are partially supported by Capes, CNPq and FAPERJ.

\end{document}